Article

# Analytical Investigations on Carrier Phase Recovery in Dispersion-Unmanaged *n*-PSK Coherent Optical Communication Systems

**Tianhua Xu**[1,2,*], **Gunnar Jacobsen**[3,4], **Sergei Popov**[4], **Jie Li**[3], **Tiegen Liu**[2], **Yimo Zhang**[2], and **Polina Bayvel**[1]

1.  University College London, London, WC1E 7JE, United Kingdom; tianhua.xu@ucl.ac.uk
2.  Tianjin University, Tianjin, 300072, China; xutianhua@tju.edu.cn
3.  Acreo Swedish ICT AB, Stockholm, SE-16425, Sweden; gunnar.jacobsen@acreo.se
4.  Royal Institute of Technology, Stockholm, SE-16440, Sweden; sergeip@kth.se
*   Correspondence: tianhua.xu@ucl.ac.uk; xutianhua@tju.edu.cn; Tel.: +44-770-966-2611

**Abstract:** Using coherent optical detection and digital signal processing, laser phase noise and equalization enhanced phase noise can be effectively mitigated using the feed-forward and feed-back carrier phase recovery approaches. In this paper, theoretical analyses of feed-back and feed-forward carrier phase recovery methods have been carried out in the long-haul high-speed *n*-level phase shift keying (*n*-PSK) optical fiber communication systems, involving a one-tap normalized least-mean-square (LMS) algorithm, a block-wise average algorithm, and a Viterbi-Viterbi algorithm. The analytical expressions for evaluating the estimated carrier phase and for predicting the bit-error-rate (BER) performance (such as the BER floors) have been presented and discussed in the *n*-PSK coherent optical transmission systems by considering both the laser phase noise and the equalization enhanced phase noise. The results indicate that the Viterbi-Viterbi carrier phase recovery algorithm outperforms the one-tap normalized LMS and the block-wise average algorithms for small phase noise variance (or effective phase noise variance), while the one-tap normalized LMS algorithm shows a better performance than the other two algorithms for large phase noise variance (or effective phase noise variance). In addition, the one-tap normalized LMS algorithm is more sensitive to the level of modulation formats.

## 1. Introduction

Since the first generation of optical fiber communication systems was deployed over 30 years ago, the achievable data rates carried by a single optical fiber have been raised over 10,000 times, and the data network traffic has also been increased by over a factor of 100 [1,2]. Up to date, more than 90% of the digital data is transmitted over optical fibers, to constitute the great part of the national and the international telecommunication infrastructures. The effective information capacity of these networks has been widely increased over the past three decades with the introduction and development of wavelength division multiplexing (WDM), higher-level modulation formats, digital signal processing (DSP), advanced optical fibers and amplification technologies [3,4]. These developments promoted the revolution of communication systems and the growth of Internet, towards the direction of higher-speed and longer-distance transmissions [2,3]. The performance of long-haul high-speed optical fiber communication systems can be significantly degraded by the impairments in the transmission channels and laser sources, such as chromatic dispersion (CD), polarization mode dispersion (PMD), laser phase noise (PN) and fiber nonlinearities (FNLs) [4-8]. Using coherent optical detection and digital signal processing, the powerful equalization and effective mitigation of the communication system impairments can be implemented in the electrical domain [9-18], which has become one of the most promising techniques for the next-generation optical fiber communication networks to achieve a performance very close to the Shannon capacity limit [19,20], with an entire capture of the amplitude and phase of the optical signals. Using high-level modulation formats such as the *n*-level phase shift keying (*n*-PSK) and the *n*-level quadrature amplitude modulation (*n*-QAM), the performance of optical fiber transmission systems will be degraded seriously by the phase noise from the transmitter (Tx) lasers and the local oscillator (LO) lasers [21,22]. To compensate the phase noise from the laser sources, some feed-forward and feed-back carrier phase recovery (CPR) approaches have been proposed to estimate and remove the phase of optical carriers [23-32]. Among these carrier phase estimation (CPE) methods, the one-tap normalized least-mean-square (LMS) algorithm, the block-wise average (BWA) algorithm, and the Viterbi-Viterbi (VV) algorithm have been validated for mitigating the laser phase noise effectively, and are also regarded as the most promising DSP algorithms in the real-time implementation of the high-speed coherent optical fiber transmission systems [27-32]. Thus it will be of importance and interest to investigate the performance of these three carrier phase recovery algorithms in long-haul high-speed optical communication systems.

In the electronic dispersion compensation (EDC) based coherent optical fiber communication systems, an effect of equalization enhanced phase noise (EEPN) will be generated due to the interactions between the electronic dispersion equalization module and the laser phase noise (in the post-EDC case the EEPN comes from the LO laser) [33-38]. The performance of long-haul optical fiber communication systems will be degraded seriously due to the equalization enhanced phase noise, with the increment of fiber dispersion, laser linewidths, modulation levels, symbol rates and system bandwidths [33-36]. The impacts of EEPN have been investigated in the single-channel, the WDM multi-channel, the orthogonal frequency division multiplexing (OFDM), the dispersion pre-distorted, and the multi-mode optical fiber transmission systems [39-46]. In addition, some investigations have been carried out to study the influence of EEPN in the carrier phase recovery in long-haul high-speed optical communication systems [47-50]. Considering the equalization enhanced phase noise, the traditional analyses of the carrier phase recovery approaches are not suitable any longer for the design and the optimization of long-haul high-speed optical fiber networks. Therefore, it will also be interesting and useful to investigate the performance of the one-tap normalized LMS, the block-wise average, and the Viterbi-Viterbi carrier phase recovery algorithms, when the influence of equalization enhanced phase noise is taken into account.

In the previous reports, the analytical derivations and the numerical studies for the one-tap normalized LMS, the block-wise average, and the Viterbi-Viterbi carrier phase recovery methods have been carried out based on the quadrature phase shift keying (QPSK) coherent optical transmission system [26,37,51]. However, with the development of the optical fiber networks and the increment of transmission data capacity, the QPSK modulation format cannot satisfy the demand of the high-speed optical fiber communication systems any more. Therefore, the analyses on the carrier phase recovery approaches should also be updated accordingly for the optical fiber transmission systems using higher-level modulation formats, such as the *n*-PSK communication systems.

In this paper, built on the previous work in Ref [26,37,51], the theoretical assessments of the carrier phase recovery using the one-tap normalized LMS, the block-wise average, and the Viterbi-Viterbi algorithms are extended and analyzed in detail for the long-haul high speed *n*-PSK coherent optical fiber communication systems, considering both the intrinsic laser phase noise and the equalization enhanced phase noise. The analytical expressions for the estimated carrier phase in the one-tap normalized LMS, the block-wise average, and the Viterbi-Viterbi algorithms has been derived, and the BER performance such as the BER floors in these three carrier phase recovery approaches has been predicted for the *n*-PSK coherent optical transmission systems. Our results indicate that the Viterbi-Viterbi carrier phase recovery algorithm outperforms the one-tap normalized LMS and the block-wise average algorithms for small phase noise variance (or effective phase noise variance), while the one-tap normalized LMS carrier phase recovery algorithm shows a better performance than the other two algorithms for large phase noise variance (or effective phase noise variance). It is also found that the one-tap normalized LMS algorithm is more sensitive to the level of the modulation formats than the other two algorithms.

**2. Laser phase noise and equalization enhanced phase noise**

As shown in Fig. 1, the origin of equalization enhanced phase noise in the coherent optical communication systems using electronic dispersion compensation and carrier phase recovery is schematically illustrated. In such systems, the transmitter laser phase noise goes through both the transmission fiber and the EDC module, and therefore the net dispersion experienced by the transmitter laser phase noise is close to zero. However, the LO laser phase noise only goes through the EDC module, where the transfer function is heavily dispersed in the transmission system without using any optical dispersion compensation (ODC) techniques. As a result, the LO laser phase noise will interplay with the dispersion equalization module, and will significantly degrade the performance of the long-haul high-speed coherent optical fiber communication systems, with the increment of fiber dispersion, laser linewidths, modulation formats, and symbol rates [33,34,36].

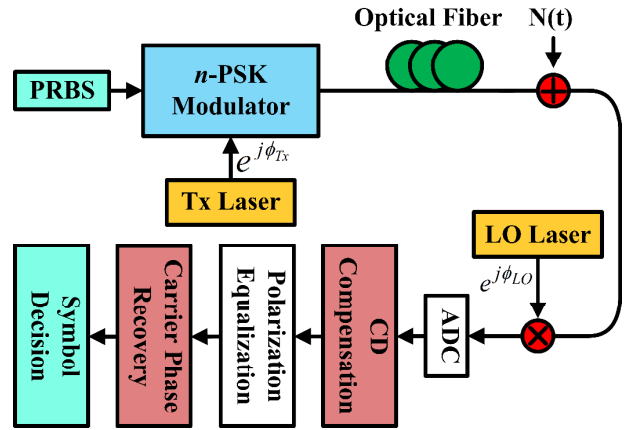

**Figure 1.** Principle of equalization enhanced phase noise in electronic dispersion compensation based *n*-PSK coherent optical transmission system. PRBS: pseudo random bit sequence, N(t): additive white Gaussian noise (AWGN) e.g. amplified spontaneous emission (ASE) noise from optical amplifiers, ADC: analog-to-digital convertor.

In coherent optical communication systems, the variance of the phase noise coming from the transmitter laser and the LO laser follows a Lorentzian distribution and can be described using the following equation [4,5]:

$$\sigma_{Laser}^2 = 2\pi(\Delta f_{Tx} + \Delta f_{LO}) \cdot T_S , \quad (1)$$

where $\Delta f_{Tx}$ and $\Delta f_{LO}$ are the 3-dB linewidths (assuming the Lorentzian distribution) of the transmitter laser and the LO laser respectively, and $T_S$ is the symbol period of the coherent transmission system. It can be found that the variance of the laser phase noise decreases with the increment of the signal symbol rate $R_S = 1/T_S$.

Considering the interplay between the electronic dispersion compensation module and the LO laser phase noise, the noise variance of the equalization enhanced phase noise in the long-haul high-speed optical fiber communication systems can be expressed as follows [33,37,40]:

$$\sigma_{EEPN}^2 = \pi\lambda^2 D \cdot L \cdot \Delta f_{LO} / 2cT_S , \quad (2)$$

where $f_{LO}$ is the central frequency of the LO laser, which is equal to the central frequency of the transmitter laser $f_{Tx}$ in the homodyne optical communication systems, $D$ is the CD coefficient of the transmission fiber, $L$ is the length of the transmission fiber, $R_S$ is the signal symbol rate of the communication system, and $\lambda = c/f_{Tx} = c/f_{LO}$ is the central wavelength of the optical carrier wave.

When the equalization enhanced phase noise is taken into account in the carrier phase recovery, the total noise variance (or effective phase noise variance) in the long-haul high-speed *n*-PSK optical fiber transmission systems can be calculated and described using the following expression [37,40]:

$$\sigma_T^2 \approx \sigma_{Laser}^2 + \sigma_{EEPN}^2 = 2\pi T_S(\Delta f_{Tx} + \Delta f_{LO}) + \pi\lambda^2 D \cdot L \cdot \Delta f_{LO} / 2cT_S \quad (3)$$

The equalization enhanced phase noise differs from the laser phase noise, and the noise variance in Eq. (2) has two-thirds contribution in the phase noise and one-third contribution in the amplitude noise [33,37,40]. Therefore, Eq. (3) is only valid for *n*-PSK communication systems, and the performance of *n*-QAM transmission systems has to be assessed based on the evaluation of error vector magnitudes [52].

Corresponding to the definition of the laser phase noise from the transmitter and the LO lasers, an effective linewidth Δ$f_{Eff}$ can be employed to describe the total phase noise variance in the EDC based *n*-PSK coherent optical communication systems and it can be expressed as follows:

$$\Delta f_{Eff} \approx \left(\sigma_{Laser}^2 + \sigma_{EEPN}^2\right)/2\pi T_S . \tag{4}$$

When the impact from laser phase noise and the influence from EEPN give an equal contribution in the *n*-PSK optical fiber communication systems, namely $\sigma_{Laser}^2 = \sigma_{EEPN}^2$, we will have the transmission distance of $L_0 = 8cT_S^2/\lambda^2 D$.

Take the 32-Gbaud *n*-PSK coherent optical fiber transmission system as an example, and we assume that the fiber CD coefficient is 17 ps/nm/km and the central wavelengths of the transmitter and the LO lasers are both 1550 nm. In this case, we have $L_0$ = 60.69 km. It means that at this transmission distance, the laser phase noise and the EEPN will have the same impact on the degradation of the performance of the 32-Gbaud *n*-PSK optical transmission systems.

## 3. Analysis of carrier phase recovery approaches

### 3.1. One-tap normalized LMS carrier phase recovery

As a feed-back carrier phase recovery approach [27,28], which is schematically shown in Fig. 2, the transfer function of the one-tap normalized LMS algorithm in the *n*-PSK coherent optical communication systems can be expressed using the following equations:

$$y(k) = w_{NLMS}(k)x(k), \tag{5}$$

$$w_{NLMS}(k+1) = w_{NLMS}(k) + \mu e(k)x^*(k)/|x(k)|^2 , \tag{6}$$

$$e(k) = d(k) - y(k), \tag{7}$$

where *x*(*k*) is the complex input symbol, *k* is the index of the symbol, *y*(*k*) is the complex output symbol, $w_{NLMS}$(*k*) is the tap weight of the one-tap normalized LMS equalizer, *d*(*k*) is the desired output symbol after the carrier phase recovery, *e*(*k*) is the estimation error between the output symbol and the desired output symbol, and *μ* is the step size of the one-tap normalized LMS algorithm.

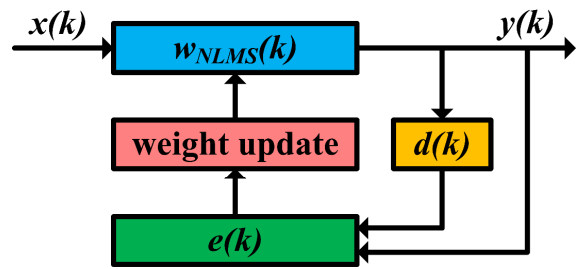

**Figure 2.** Schematic of one-tap normalized LMS carrier phase recovery algorithm.

It has been verified that the optimized one-tap normalized LMS carrier phase recovery in the QPSK optical transmission systems behaves similar to the ideal differential carrier phase recovery [24,37], and the BER floor of the one-tap normalized LMS carrier phase recovery in the QPSK transmission systems can be approximately described as follows [37]:

$$BER_{floor}^{NLMS\_QPSK} \approx \frac{1}{2} erfc\left(\frac{\pi}{4\sqrt{2}\sigma_T}\right). \tag{8}$$

Therefore, the BER floor of the one-tap normalized LMS carrier phase recovery for the *n*-PSK optical fiber communication systems can be derived accordingly, and can be expressed using the following equation:

$$BER_{floor}^{NLMS} \approx \frac{1}{\log_2 n} erfc\left(\frac{\pi}{n\sqrt{2}\sigma_T}\right), \quad (9)$$

where $\sigma_T^2$ is the total phase noise variance (or effective phase noise variance) in the long-haul high-speed *n*-PSK optical transmission systems.

*3.2. Block-wise average carrier phase recovery*

As an *n*-th power feed-forward carrier phase recovery approach, which is schematically shown in Fig. 3, the block-wise average algorithm calculates the *n*-th power of the received symbols to remove the information of the modulated phase in the *n*-PSK coherent transmission systems, and the computed phase (*n*-th power) are summed and averaged over a certain block. The averaged phase value is then divided over *n*, and the final result is regarded as the estimated phase for the received symbols within the entire block [29,30]. For the *n*-PSK coherent optical communication systems, the estimated carrier phase for each process block using the block-wise average algorithm can be expressed as:

$$\phi_{BWA}(k) = \frac{1}{n}\arg\left\{\sum_{p=1+(q-1)\cdot N_{BWA}}^{q\cdot N_{BWA}} x^n(p)\right\}, \quad (10)$$

$$q = \lceil k/N_{BWA} \rceil, \quad (11)$$

where *k* is the index of the received symbol, $N_{BWA}$ is the block length in the block-wise average algorithm, and $\lceil x \rceil$ means the closest integer lager than *x*.

The BER floor of the block-wise average carrier phase recovery in the *n*-PSK coherent optical communication systems can be derived using the Taylor series expansion, and can be approximately described using the following equation:

$$BER_{floor}^{BWA} \approx \frac{1}{N_{BWA}\log_2 n}\cdot\sum_{p=1}^{N_{BWA}} erfc\left(\frac{\pi}{n\sqrt{2}\sigma_{BWA}(p)}\right), \quad (12)$$

$$\sigma_{BWA}^2(p) = \frac{\sigma_T^2\left[2(p-1)^3 + 3(p-1)^2 + 2(N_{BWA}-p)^3 + 3(N_{BWA}-p)^2 + N_{BWA}-1\right]}{6N_{BWA}^2}, \quad (13)$$

where $\sigma_T^2$ is the total phase noise variance (or effective phase noise variance) in the long-haul high-speed *n*-PSK optical transmission systems.

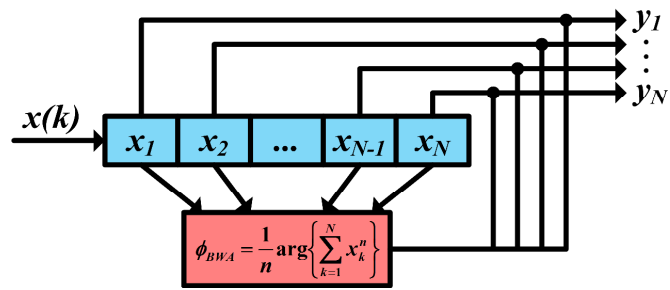

**Figure 3.** Schematic of block-wise average carrier phase recovery algorithm.

*3.3. Viterbi-Viterbi carrier phase recovery*

As another *n*-th power feed-forward carrier phase recovery approach, which is schematically shown in Fig. 4, the Viterbi-Viterbi algorithm also calculates the *n*-th power of the received symbols to remove the information of the modulated phase. The computed phase are also summed and averaged over the processing block (with a certain block length). Compared to the block-wise average algorithm, the Viterbi-Viterbi algorithm just treats the extracted phase as the estimated phase for the central symbol in each processing block [31,32]. The estimated carrier phase in the Viterbi-Viterbi algorithm in the *n*-PSK coherent optical transmission systems can be described using the following equation:

$$\phi_{VV}(k) = \frac{1}{n}\arg\left\{\sum_{q=-(N_{VV}-1)/2}^{(N_{VV}-1)/2} x^n(k+q)\right\}, \tag{14}$$

where $N_{VV}$ is the block length of the Viterbi-Viterbi algorithm, and should be an odd value of e.g. 1,3,5,7…

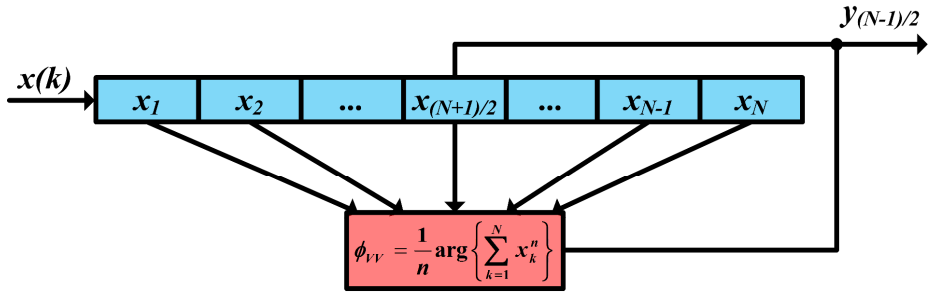

**Figure 4.** Schematic of Viterbi-Viterbi carrier phase recovery algorithm.

Using the Taylor expansion, the BER floor of the Viterbi-Viterbi carrier phase recovery in the *n*-PSK coherent optical communication systems can be assessed analytically, and can be expressed approximately using the following equation:

$$BER_{floor}^{VV} \approx \frac{1}{\log_2 n} erfc\left(\frac{\pi}{n \cdot \sqrt{\frac{N_{VV}^2 - 1}{6 N_{VV}}} \cdot \sigma_T}\right), \tag{15}$$

where $\sigma_T^2$ is the variance of the total phase noise (or effective phase noise) in the long-haul high-speed *n*-PSK optical fiber transmission systems.

## 4. Results and Discussions

*4.1. Results*

In this section, the performance of carrier phase recovery in the long-haul high-speed *n*-PSK coherent optical transmission systems using the one-tap normalized LMS, the block-wise average, and the Viterbi-Viterbi algorithms are investigated based on the above theoretical studies, which can be regarded as the extensions of the QPSK results in the previous reported work [26,37,51]. The comparisons of the three carrier phase recovery approaches have also been carried out in detail. In all these analyses and discussions, the standard single mode fiber (SSMF) has been employed in the *n*-PSK coherent optical transmission systems, where the fiber dispersion is 17 ps/nm/km, the central wavelengths of both the transmitter laser and the LO laser are 1550 nm, and the signal symbol rate is 32-Gbaud. The fiber attenuation, the PMD, and the nonlinear effects are all neglected.

Based on Eq. (9), the performance of the one-tap normalized LMS carrier phase recovery in the coherent optical fiber communication systems using different modulation formats is shown in Fig. 5. The optimization of the one-tap normalized LMS carrier phase recovery has been investigated and discussed in detail in Ref [27,37], where the step size varying from 0.01 to 1 has been applied. A smaller step size will degrade the BER floor due to the fast phase changing occurring in the long effective symbol average-span, while a larger step size will degrade the signal-to-noise ratio (SNR) sensitivity in the one-tap normalized LMS carrier phase recovery. The optimal step sizes for different effective linewidths have been studied, with a resolution of 0.005 used in the optimization process [37]. In this section we assume that all the operations of the one-tap normalized LMS algorithm have been optimized. It can be found in Fig. 5 that the one-tap normalized LMS carrier phase recovery algorithm is very sensitive to the phase noise variance and the modulation formats, especially when phase noise variance is less than 0.1.

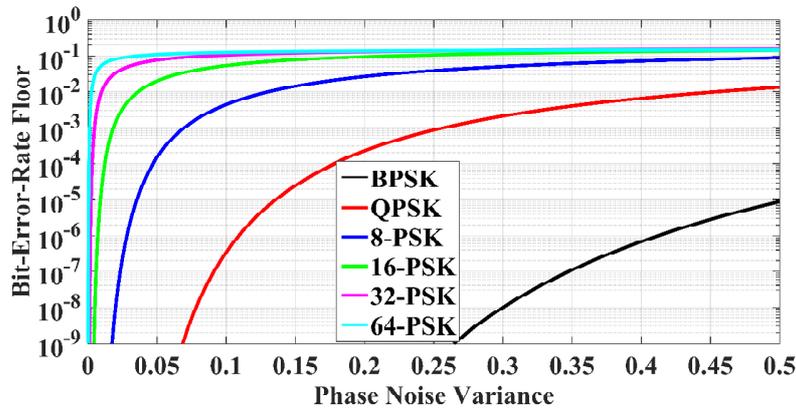

**Figure 5.** BER floors versus phase noise variance in the one-tap normalized LMS carrier phase recovery in the coherent optical transmission systems using different modulation formats.

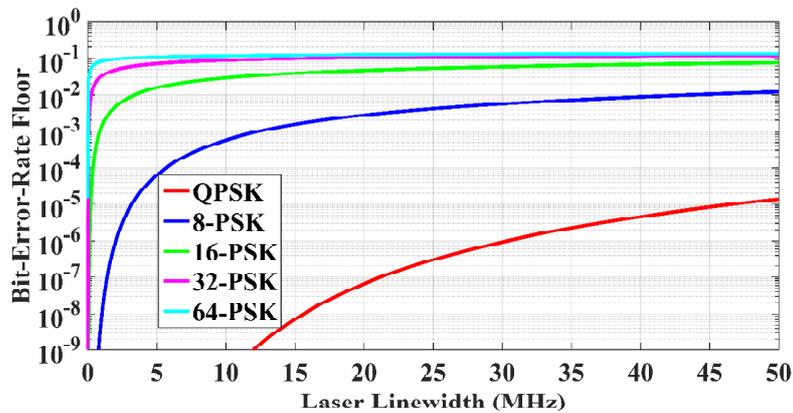

**Figure 6.** BER floors versus laser linewidths in the one-tap normalized LMS carrier phase recovery in the optical fiber transmission systems using different modulation formats. The indicated linewidth value is the 3-dB linewidth for both the Tx and the LO lasers.

In the case of back-to-back (BtB) or without considering EEPN, the performance of BER floors versus laser linewidths in the one-tap normalized LMS carrier phase estimation has been studied in Fig. 6 based on the analysis in Eq. (1) and Eq. (9). The indicated linewidth value in Fig. 6 is the 3-dB linewidth for both the transmitter laser and the LO laser. It can be seen that the BER floors of the one-tap normalized LMS carrier phase recovery are deteriorated with the increment of the modulation format levels and the laser linewidths. The degradations due to the laser phase noise (laser linewidths) are more drastic for higher-level modulation formats.

Considering the impact of EEPN, the effective phase noise variance will increase with the increment of the transmission distance and the laser linewidth. As shown in Fig. 7, the BER floors of the one-tap normalized LMS carrier phase recovery has been investigated for different transmission distances, when the linewidths of both the Tx and the LO lasers are set to 1 MHz. It can be found that the performance of the one-tap normalized LMS carrier phase recovery is degraded significantly with the increment of transmission distances, and this effect is more severe for higher-level modulation formats due to the less tolerance to laser phase noise and EEPN.

The performance of the block-wise average carrier phase recovery approach has also been investigated, as shown in Fig. 8, Fig. 9 and Fig. 10. The BER floors versus different phase noise variances (or effective phase noise variance) in the block-wise average carrier phase recovery is described in Fig. 8. In Fig. 8(a), the performance of the block-wise average algorithm is studied in terms of different block lengths in the 8-PSK optical transmission system. It can be found that the phase noise (or effective phase noise) induced BER floor in the block-wise average carrier phase recovery algorithm is increased with the increment of block length. Generally, a smaller block length will lead to a lower phase noise induced BER floor due to a more accurate estimation of carrier phase, while a larger block length is more effective for mitigating the amplitude noise (such as ASE noise) to improve the SNR sensitivity. In practical transmission systems, the optimal block length is determined by considering the trade-off between the phase noise and the amplitude noise. As an example, the block length of $N_{BWA}$ = 11 is employed in all the subsequent analyses, if the value is not specified. Based on Eq. (12) and Eq. (13), the performance of the block-wise average carrier phase recovery in the coherent optical communication systems using different modulation formats is shown in Fig. 8(b), where the block length is 11. It can be found in Fig. 8(b) that the block-wise average carrier phase recovery algorithm is also very sensitive to the phase noise variance and the modulation formats, when phase noise variance is less than 0.1.

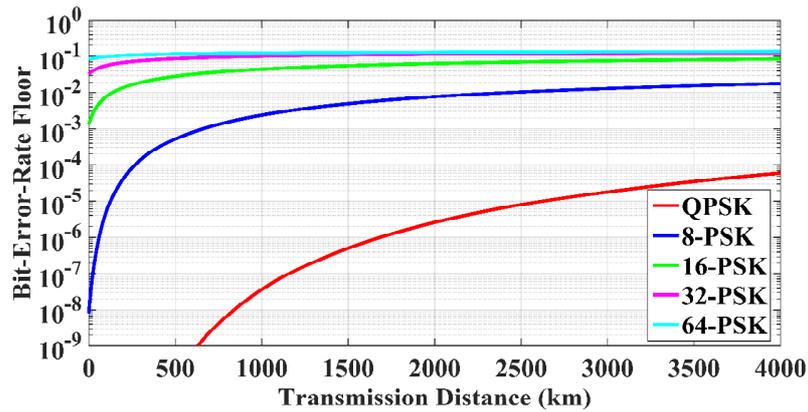

**Figure 7.** BER floors versus transmission distances in the one-tap normalized LMS carrier phase recovery in the coherent optical transmission systems using different modulation formats, considering the equalization enhanced phase noise. Both the Tx and LO lasers linewidths are 1 MHz.

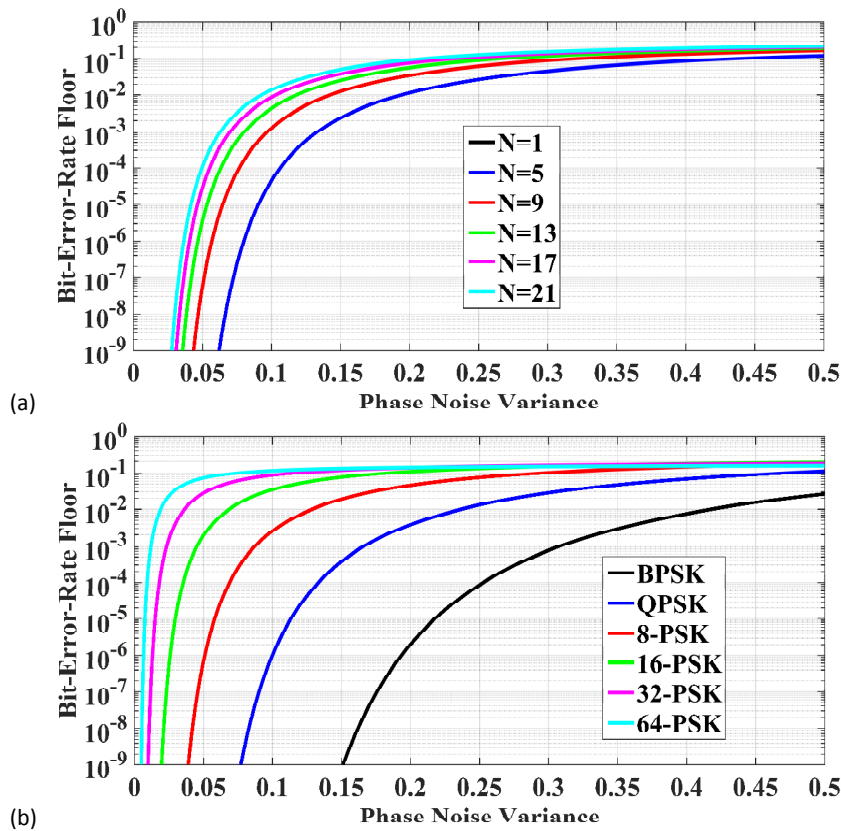

**Figure 8.** BER floors versus phase noise variances in the block-wise average carrier phase recovery in the coherent optical transmission systems. (a) different block lengths in the 8-PSK transmission system, (b) different modulation formats with the block length of 11.

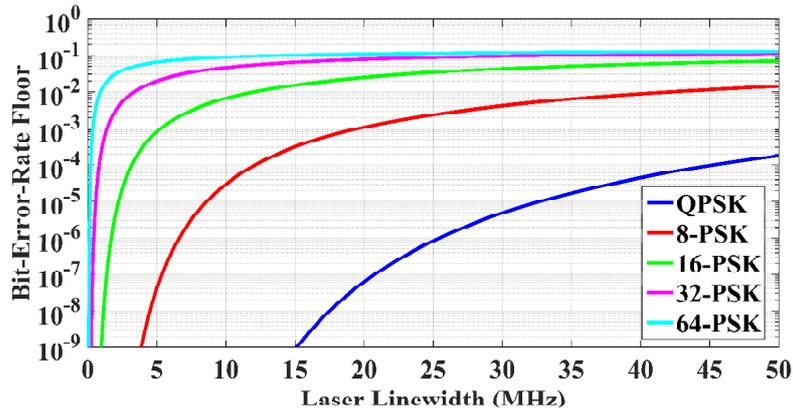

**Figure 9.** BER floors versus laser linewidths in the block-wise average carrier phase recovery in the coherent optical transmission systems using different modulation formats. The block length is 11, and the indicated linewidth value is the 3-dB linewidth for both the Tx and the LO lasers.

Based on the analyses in Eq. (1), Eq. (12) and Eq. (13), the BER floors versus laser linewidths in the block-wise average carrier phase recovery for the back-to-back case or without considering EEPN has been studied in Fig. 9, where the indicated linewidth value is the 3-dB linewidth for both the transmitter laser and the LO laser. It can be found that the BER floors in the block-wise average carrier phase recovery are also degraded with the increment of modulation format levels and laser linewidths. The degradations due to the laser phase noise (laser linewidths) are also more severe for higher-level modulation formats.

As shown in Fig. 10, the BER floors of the block-wise average carrier phase recovery have been investigated for different transmission distances by considering the impact of EEPN, where the linewidths of both the Tx and the LO lasers are set to 1 MHz. It can be found that the performance of the block-wise average carrier phase recovery is degraded significantly with the increment of transmission distances, and this effect is more severe for higher-level modulation formats, since the EEPN influence will scale with the increment of transmission distances and modulation formats.

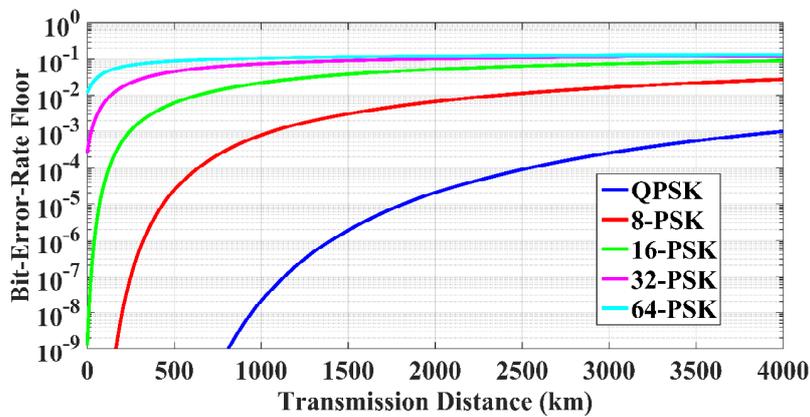

**Figure 10.** BER floors versus transmission distances in the block-wise average carrier phase recovery in the coherent optical transmission systems using different modulation formats. The block length is 11, and the linewidth of both the Tx and the LO lasers are 1MHz.

From Fig. 11 to Fig. 13, the performance of the Viterbi-Viterbi carrier phase recovery approach has been investigated in terms of the phase noise variances, the laser linewidths and the transmission distances. The BER floors versus the phase noise variances in the Viterbi-Viterbi carrier phase recovery algorithm is studied in Fig. 11. In Fig. 11(a), the BER floors are also studied in terms of different block lengths in the Viterbi-Viterbi CPR algorithm for the 8-PSK optical transmission system. Similar to the block-wise average algorithm, a smaller block length in the Viterbi-Viterbi carrier phase recovery will generate a lower phase noise induced BER floor, in contrast, a larger block length is more tolerant to the amplitude noise and will lead to a better SNR sensitivity. The optimal block length is again determined by considering the trade-off between phase noise and amplitude noise. It can be found that the phase noise induced BER floors in the Viterbi-Viterbi carrier phase recovery algorithm are also deteriorated with the increment of block lengths. Similar to the block-wise average algorithm, the block length of $N_{VV}$ = 11 is also selected as an example in the Viterbi-Viterbi carrier phase recovery to consider the mitigation of both the phase noise and the amplitude noise in practical applications. Based on Eq. (15), the performance of the Viterbi-Viterbi carrier phase recovery in the coherent optical communication systems using different modulation formats is shown in Fig. 11(b), where the block length is 11. It is found in Fig. 11(b) that the Viterbi-Viterbi carrier phase recovery algorithm is also very sensitive to the phase noise variance and the modulation formats, when phase noise variance is less than 0.15.

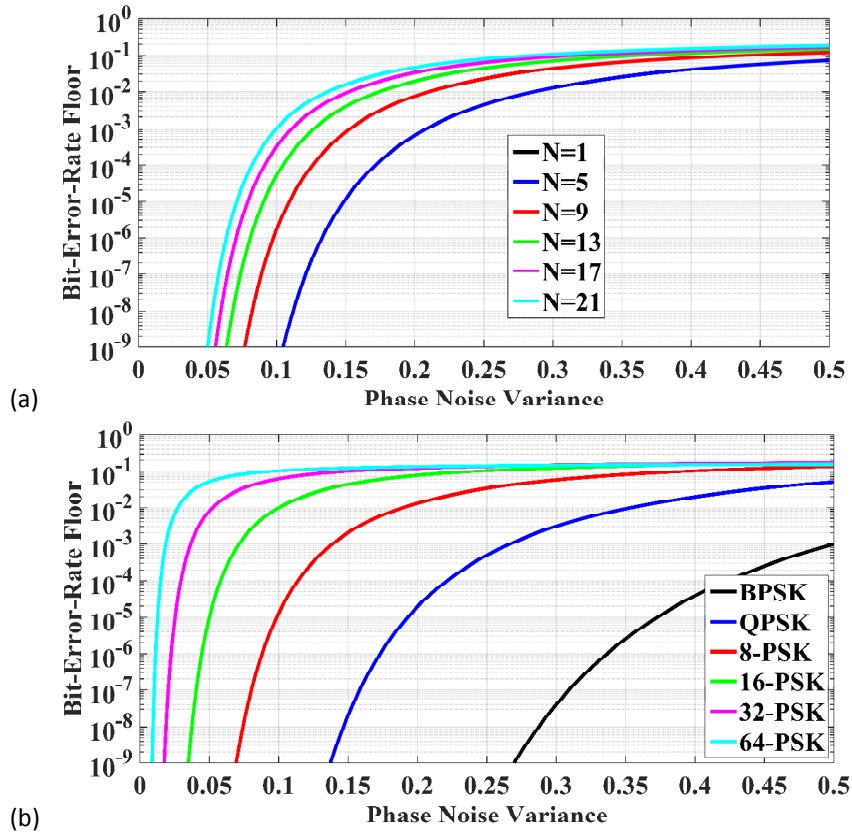

**Figure 11.** BER floors versus phase noise variances in the Viterbi-Viterbi carrier phase recovery in the coherent optical transmission systems. (a) different block lengths in the 8-PSK transmission system, (b) different modulation formats with the block length of 11.

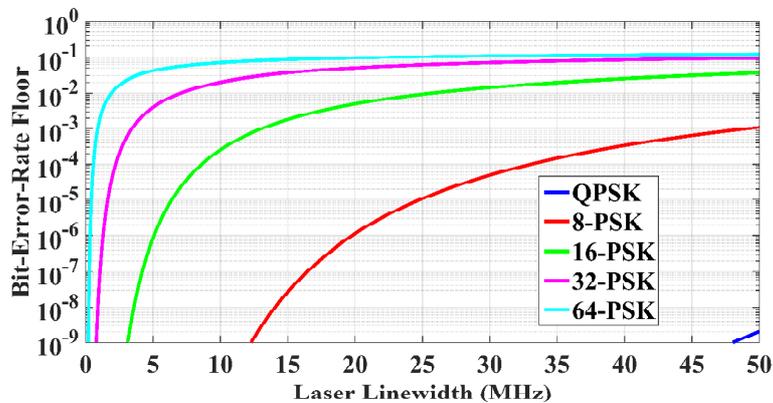

**Figure 12.** BER floors versus laser linewidths in the Viterbi-Viterbi carrier phase recovery in the coherent optical transmission systems using different modulation formats. The block length is 11, and the indicated linewidth value is the 3-dB linewidth for both the Tx and the LO lasers.

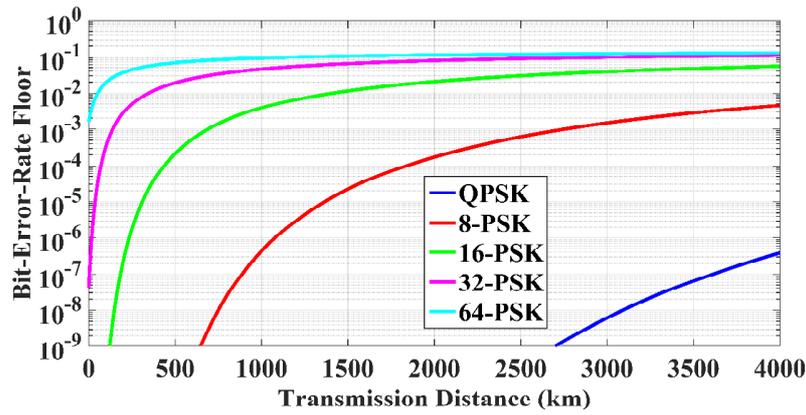

**Figure 13.** BER floors versus transmission distances in the Viterbi-Viterbi carrier phase recovery in the coherent optical transmission systems using different modulation formats. The block length is 11, and the linewidth of both the Tx and the LO lasers are 1MHz.

Without considering EEPN (or for the back-to-back case), the BER floors versus laser linewidths in the Viterbi-Viterbi carrier phase recovery have been studied in Fig. 12 based on the analyses in Eq. (1) and Eq. (15), where the indicated linewidth value is again the 3-dB linewidth for both the transmitter laser and the LO laser. It can be found that the BER floors are also degraded significantly with the increment of laser linewidths, and this degradation is also more severe with the increment of modulation format levels.

As shown in Fig. 13, the BER floors of the Viterbi-Viterbi carrier phase recovery have also been investigated for different transmission distances considering the impact of EEPN, where the linewidths of the Tx and the LO lasers are both set to 1 MHz. It can be found that in the Viterbi-Viterbi carrier phase recovery, the EEPN influence will also increase with the increment of transmission distances and modulation formats. The performance of the Viterbi-Viterbi algorithm is degraded significantly with the increment of transmission distances, and this effect is more serious for higher-level modulation formats.

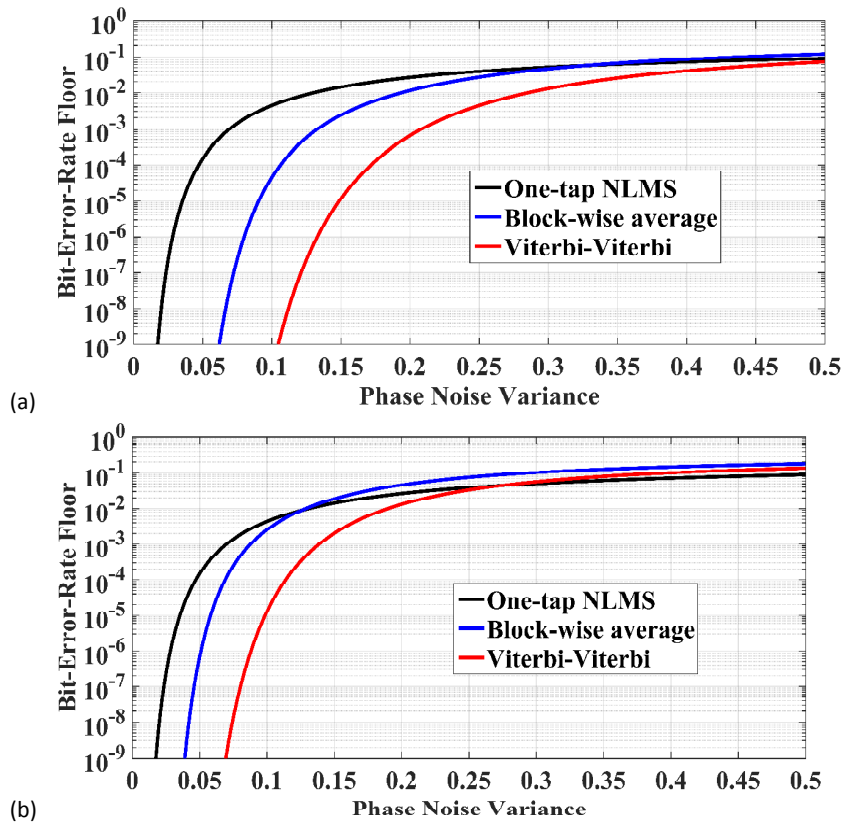

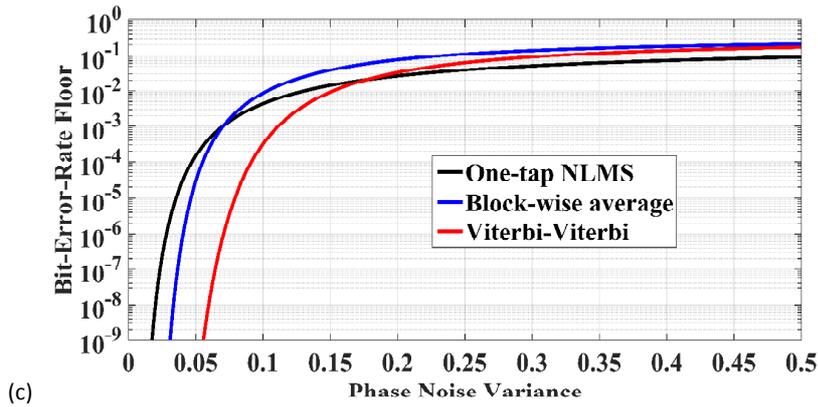

**Figure 14.** BER floors versus different phase noise variances in the three carrier phase recovery algorithms in the 8-PSK optical fiber communication systems. (a) block length of the BWA and VV algorithms is 5, (b) block length of the BWA and VV algorithms is 11, (c) block length of the BWA and VV algorithms is 17.

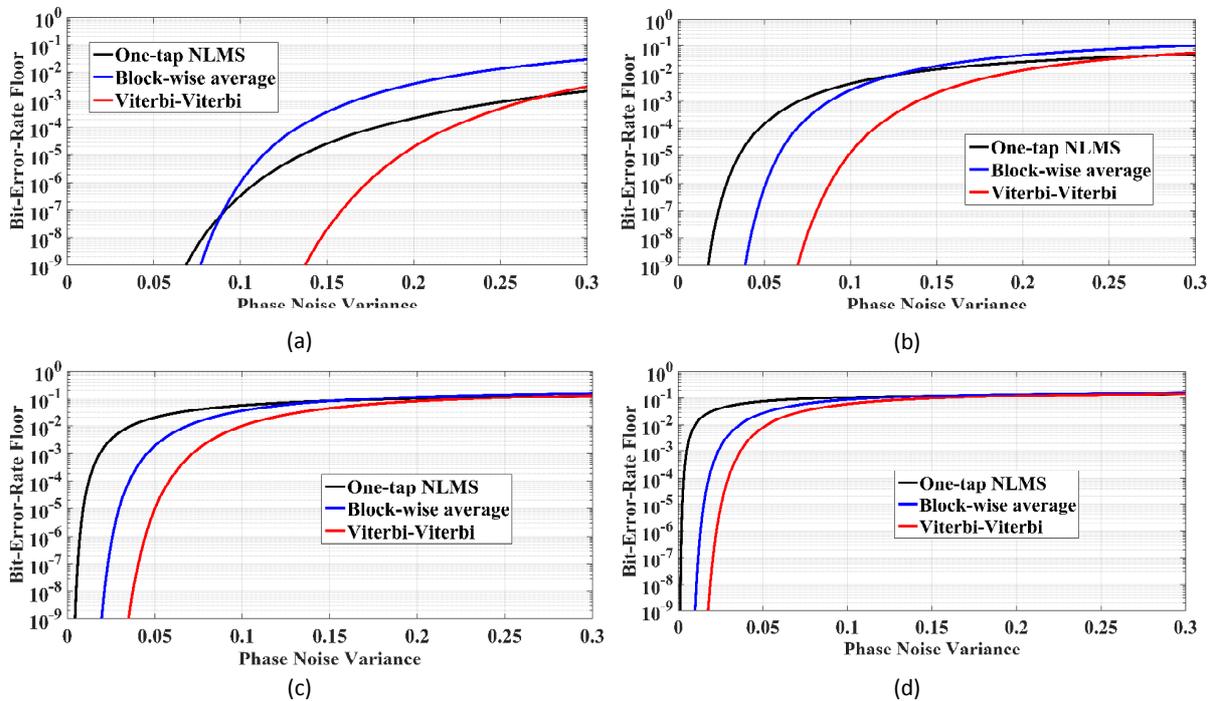

**Figure 15.** BER floors versus different phase noise variances in the three carrier phase recovery algorithms in the optical fiber communication systems using different modulation formats. Block lengths of the BWA and VV algorithms are both 11. (a) QPSK system, (b) 8-PSK system, (c) 16-PSK system, (d) 32-PSK system.

The comparisons of the one-tap normalized LMS, the block-wise average, the Viterbi-Viterbi carrier phase recovery algorithms have also been investigated in detail. The BER floors versus different phase noise variances in the above three carrier phase recovery algorithms in the 8-PSK optical fiber communication system have been studied and are shown in Fig. 14, where the block length in the block-wise average and the Viterbi-Viterbi algorithms varies from 5 to 17, in Fig. 14(a), Fig. 14(b) and Fig. 14(c), respectively. It can be seen that the phase noise induced BER floors in the block-wise average and Viterbi-Viterbi algorithms are degraded with increment of the block length, and the one-tap normalized LMS algorithm keeps the same performance due to its optimized operation. It is also found in Fig. 14 that for the small phase noise variance (or effective phase noise variance), the Viterbi-Viterbi carrier phase recovery algorithm outperforms the one-tap normalized LMS and the block-wise average algorithms, while for the large phase noise variance (or effective phase noise variance), the one-tap normalized LMS algorithm shows a better performance than the other two algorithms in the carrier phase recovery.

As shown in Fig. 15 and Fig.16, the comparison of the one-tap normalized LMS, the block-wise average, the Viterbi-Viterbi carrier phase recovery algorithms has also been investigated in terms of different modulation formats. Here a block length of 11 is used in the block-wise average and the Viterbi-Viterbi algorithms. It can be found in Fig. 15 that the BER floors in all the three algorithms are increased with the increment of modulation formats, but the variation in the one-tap normalized LMS CPR algorithm is larger than in the other two algorithms. Therefore, the one-tap normalized LMS algorithm is more sensitive to the level of the modulation formats.

Figure 16 shows the performance of BER floors versus transmission distances in the three carrier phase recovery methods under different modulation formats, where the linewidths of both the transmitter and the LO lasers are set as 1 MHz and the transmission distance are set from 0 km to 5000 km. It can also be found in Fig. 16 that, similar to Fig. 15, the one-tap normalized LMS algorithm is more sensitive to the level of the modulation formats than the block-wise average and the Viterbi-Viterbi CPR algorithms. In addition, the Viterbi-Viterbi algorithm outperforms the other two carrier phase recovery approaches when the transmission distance varies from 0 km to 5000 km, while the differences between the three carrier phase recovery algorithms become smaller with the increment of modulation formats.

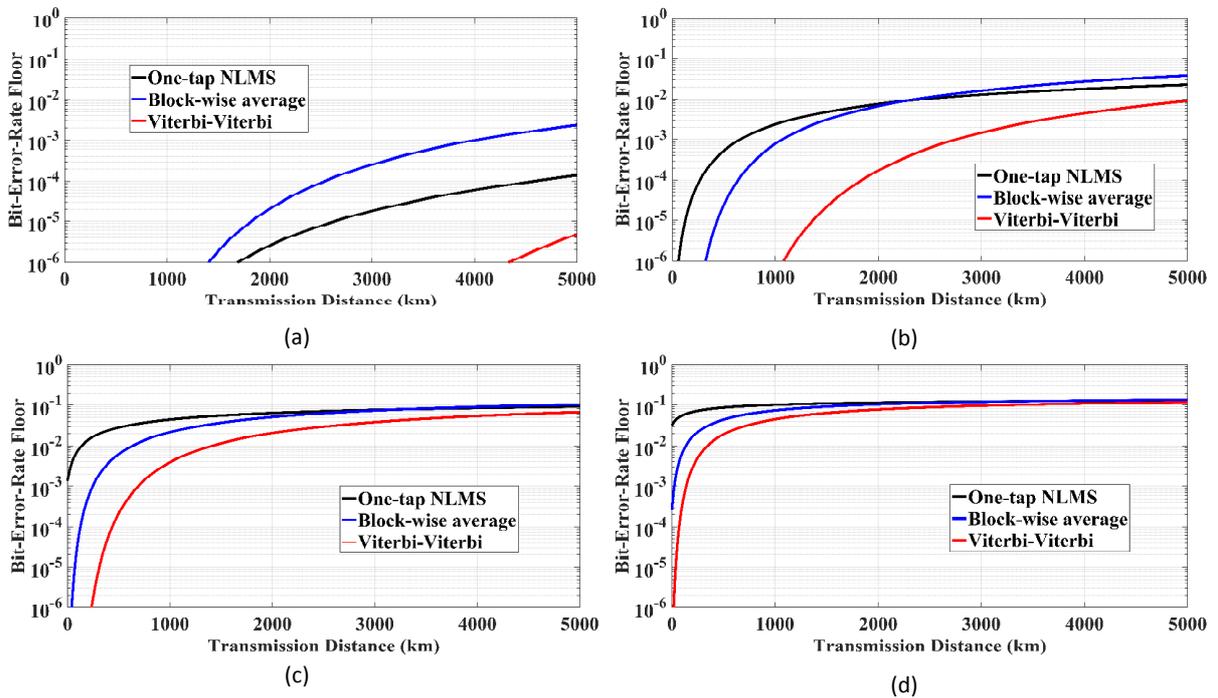

**Figure 16.** BER floors versus different transmission distances in the three carrier phase recovery algorithms in the optical fiber communication systems using different modulation formats. Linewidths of the transmitter and LO lasers are both 1 MHz, and block lengths of the BWA and VV CPR algorithms are both 11. (a) QPSK system, (b) 8-PSK system, (c) 16-PSK system, (d) 32-PSK system.

*4.2. Ideal spectral efficiency in carrier phase recovery*

For the binary symmetric channel (binary input and output alphabets, symmetric transition probability), the coding rate $R_C$ in the $n$-PSK optical fiber communication systems (assuming an ideal hard-decision forward error correction coding) can be expressed as follows [53],

$$R_C = 1 + BER \cdot \log_2(BER) + (1-BER) \cdot \log_2(1-BER). \tag{16}$$

Correspondingly, the ideal spectral efficiency (assuming an ideal hard-decision forward error correction coding) in the carrier phase recovery in the $n$-PSK coherent optical fiber communication systems can be calculated as:

$$SE = R_C \cdot N_p \cdot \log_2(n), \tag{17}$$

where $SE$ is the ideal spectral efficiency, $n$ is the modulation format level of the communication system, $N_p$ is the number of polarization states. The BER limits in Eq. (16) can be obtained from the BER floors in the three carrier phase recovery approaches according to Eq. (9), Eq. (12), and Eq. (15), respectively.

*4.3. Complexity of carrier phase recovery approaches*

The computational complexity is always a significant consideration and criterion for the DSP algorithms. Here the complexity of the one-tap normalized LMS, the block-wise average, and the Viterbi-Viterbi carrier phase recovery algorithms has been investigated in terms of the number of the complex multiplications per recovered symbol, which is shown in Table 1 ($n$ is the level of modulation formats). It is found that the computational complexity of the one-tap normalized LMS algorithm is independent from the modulation formats, while the complexity of the block-wise average and the Viterbi-Viterbi CPR algorithms scales linearly with the level of modulation formats. Note that the computations in the pre-convergence of the one-tap normalized LMS algorithm also has to be considered in the practical applications.

**Table 1.** Complexity of carrier phase recovery approaches (complex multiplications per symbol).

| One-tap normalized LMS | Block-wise average | Viterbi-Viterbi |
|---|---|---|
| 5 | $n$ | $n$ |

All the above analyses are based on the carrier phase recovery in the *n*-PSK coherent optical communication systems, however, all these discussions can be directly extended into the circular *n*-QAM transmission systems. Meanwhile, although the *n*-PSK signals have a lower tolerance to the ASE noise than the multi-amplitude signals (such as *n*-QAM signals), the *n*-PSK signals will have a better tolerance to fiber nonlinearities due to the constant amplitudes [54,55].

**5. Conclusions**

Theoretical analyses of the carrier phase recovery in long-haul high-speed *n*-PSK coherent optical fiber communication systems, using the one-tap normalized LMS, the block-wise average, and the Viterbi-Viterbi algorithms, have been investigated and described in detail, considering both the laser phase noise and the equalization enhanced phase noise. The expressions for the estimated carrier phase in these three algorithms have been presented, and the BER performance such as the BER floors, has been predicted analytically. Comparative studies of the one-tap normalized LMS, the block-wise average, and the Viterbi-Viterbi algorithms have also been carried out. It has been found that the Viterbi-Viterbi carrier phase recovery algorithm outperforms the one-tap normalized LMS and the block-wise average algorithms for small phase noise variance (or effective phase noise variance), while the one-tap normalized LMS algorithm shows a better performance than the other two algorithms in the carrier phase recovery for large phase noise variance (or effective phase noise variance). In addition, the one-tap normalized LMS carrier phase recovery algorithm is more sensitive to the level of modulation formats than the other two algorithms.

The BER floors in this paper were discussed and analyzed based on the influence from laser phase noise and equalization enhanced phase noise in the long-haul *n*-PSK transmission systems, and this represents the system limits from laser phase noise and equalization enhanced phase noise. In addition, signal degradation from fiber nonlinearities is also a significant effect in such communication systems. Therefore, the actual BER floors will be determined by involving the impacts from ASE noise, laser phase noise, equalization enhanced phase noise and fiber nonlinearities, which will be investigated in our future work.

**Acknowledgments:** This work is supported in part by UK Engineering and Physical Sciences Research Council (project UNLOC EP/J017582/1), in part by European Commission Research Council FP7-PEOPLE-2012-IAPP (project GRIFFON, No. 324391), in part by European Commission Research Council FP7-PEOPLE-2013-ITN (project ICONE, No. 608099), and in part by Swedish Research Council Vetenskapsradet (No. 0379801).

**Author Contributions:** T.X. presented the basic idea and carried out the analytical calculations and discussions. G.J., S.P., J.L., T.L., Y.Z. and P.B. contributed to developing the research ideas and were involved in the discussion of results. T.X. wrote the main manuscript and prepared the figures. All authors reviewed the manuscript and gave the final approval for publication.

**Conflicts of Interest:** The authors declare no conflict of interest.